\documentclass[iopscience,apj]{emulateapj}

\def\I{{\sc i}}
\def\II{{\sc ii}}
\def\III{{\sc iii}}

\newcommand{\vsini}{\ensuremath{v_{{\mathrm{e}}}\sin i}}
\def\Rh{\rule{20.0pt}{0.0pt}}

\shorttitle{Heavy elements in Sirius}  
\shortauthors{Cowley, et al.}

\begin{document}


\title{A study of the elements copper through uranium in Sirius A: \\
Contributions from STIS and ground-based spectra}

\author{C. R. Cowley}
\affil{Department of Astronomy, University of Michigan,
Ann Arbor, MI 48109-1107, USA}


\email{cowley@umich.edu} 
\author{T. R. Ayres}
\affil{Center for Astrophysics and Space Astronomy, University of
Colorado, Boulder, CO 80309-0389, USA}
\author{F. Castelli}
\affil{Instituto Nazionale di Astrofisica, Osservatorio Astronomico
di Trieste, Via Tiepolo 11,I34143 Trieste, Italy}
\author{A. F. Gulliver}
\affil{Department of Physics and Astronomy, Brandon University,
Brandon MB RTA 6A9, Canada}
\author{R. Monier}
\affil{LESIA, UMR 8109, Observatoire de Paris, Place J. Janssen,
92195 Meudon, France}
\and
\author{G. M. Wahlgren}
\affil{CSRA/STScI, 3700 San Martin Drive,
Baltimore, MD 21218, USA}


\begin{abstract}

We determine abundances or upper limits for 
all of the 55 stable elements
from copper to uranium for the A1 Vm star Sirius.  The purpose of
the study is to assemble the most complete picture of elemental
abundances with the hope of revealing the chemical history
of the brightest star in the sky, apart from the Sun.  We also
explore the relationship of this hot 
metallic-line (Am) star to its cooler 
congeners, as well as the hotter, weakly- or non-magnetic 
mercury-manganese (HgMn) stars.  
Our primary observational 
material consists of {\em Hubble Space Telescope} ($HST$) spectra taken with 
the Space Telescope Imaging Spectrograph (STIS) in the ASTRAL project.
We have also used archival material from the   
$COPERNICUS$ satellite, and from the $HST$ 
Goddard High-Resolution Spectrograph (GHRS), as well as   
ground-based spectra from Furenlid, Westin, Kurucz, Wahlgren, and their coworkers,
ESO spectra from the UVESPOP project, and NARVAL spectra retrieved from
PolarBase.  Our analysis has been primarily by spectral synthesis,
and in this work we have had the great advantage of extensive atomic
data unavailable to earlier workers.
We find most abundances as well as upper limits range from 10 to 100 
times above solar values.  We see no indication of the huge abundance
excesses of 1000 or more that occur among many chemically
peculiar (CP) stars of the upper main sequence. The picture
of Sirius as a hot Am star is reinforced.
 

\end{abstract}


\section{Introduction \label{sec:intro}}

The history of matter is written in its 
abundance patterns--a plot of abundances versus
mass number $A$, or atomic number $Z$.  
For example, the imprint of major
nucleosynthetic processes is seen in the iron peak,
and the local maxima due to the
$r-$ and $s-$ (rapid and slow) processes.
Throughout an abundance plot,
the persistent odd-even alternation of abundances,
is the imprint of nuclear processing.

The abundance patterns of stars more massive than the Sun have             
revealed a history of significant non-nuclear processes.  An
overall pattern discussed in the pioneering diffusion 
work of \citet{mic70}
was a deficiency of typically abundant elements, such as 
helium (He, $Z$ = 2)\footnote{The atomic numbers, $Z$, and symbols 
for chemical elements 
are given when the elements are first named.  Names of elements are
also given explicitly in cases where clarity is desirable.},
carbon (C, $Z$ = 6), and oxygen (O, $Z$ = 8), 
with an excess of rare species, for 
example, beyond the iron peak--gallium (Ga, $Z$ = 31), 
strontium (Sr, $Z$ = 38), yttrium (Y, $Z$ = 39),
or zirconium (Zr, $Z$ = 40).  Additionally, violations of the ubiquitous odd-even
effect were found in the Sr-Y-Zr triplet,
and even on the iron peak, where manganese (Mn, $Z$ = 25) could be more 
abundant than iron (Fe, $Z$ = 26) or chromium (Cr, $Z$ = 24) (cf. \citet{ade06}).

More recently, fascinating abundance patterns apparently 
unrelated to nucleosynthesis or diffusion have been noted in
such disparate objects as the 
$\lambda$ Bootis stars \citep{ven90}, Herbig Ae \citep{cow10,fol12}, 
and post-AGB stars \citep{van03}. These objects sometimes
show solar abundances of the lighter elements, with deficiencies
of heavier ones.     

For Sirius, as for most A-stars, abundances are 
generally known for most of the elements
through zinc (Zn, $Z$ = 30).  Beyond Zn, 
``islands of detectability'' are
found, for example at Sr, Y,
and Zr, and at barium (Ba, $Z$ = 56) and some of the 
lighter lanthanide rare earths 
(e.g. lanthanium (La, $Z$ = 57) -- gadolinium (Gd, $Z$ = 64)).
This 
leaves much of the periodic table unexplored.  Yet significant
clues to the chemical history of stars like Sirius may lie in 
these poorly explored regions.  

Lacuna in the abundance patterns occur for several reasons. 
Cosmic abundances generally decline with increasing $Z$, 
so we are less likely to see lines from heavier elements
than from lighter ones.  This trend may be compensated, when
an atomic or ionic spectrum is simple, with
most of the line strength concentrated in a few
wavelengths.  This is one reason for the islands of 
detectability.  But the strongest
lines of an atom or ion may fall in wavelength regions that
are not commonly available.  This happens for
many of the elements in the right-hand part of the periodic
table. Many strong lines fall in the satellite ultraviolet.
While the {\em International Ultraviolet Explorer (IUE)} 
spectra provided coverage of much of this region
for many stars, its noise and spectral resolution left much
room for improvement.

One of the goals of the ASTRAL Cool Star Project \citep{ayr10} was
to provide high-quality spectra in the satellite ultraviolet from
the STIS instrument of the $HST$.  
The current study is a part of an extension of that project to 
hotter stars.

\section{Resume of past work}
\label{sec:past}
We build upon a significant body of work.  \citet{all42} 
published curve of growth studies of abundances in both Sirius and
$\gamma$ Geminorum, presaging the larger work of 
\citet{koh64,ko64b}, based on numerous high-dispersion photographic
spectra.    
Information for some 3650 lines was presented in the
second reference, including a number of relevance to the current
paper.  In the English abstract of the first of these references, Kohl 
states that ``Sirius tends to be a metallic line star.''  
\citet{con65} gives references to several early works
that called attention to the similarity of Sirius's abundances
to those of Am stars.  \citet{hil03} compared abundances from 
spectral synthesis of high-resolution spectra of six superficially
normal A-stars, including Sirius.   

High-quality wavelength measurements
were not made for Sirius in earlier work. Such measurements often 
indicate the relative contributions of blends in a simple way 
that is possible, but sometimes tedious 
from a detailed synthesis, with multiple identifications.  
Accurate stellar wavelengths are not a substitute for the 
impressive techniques typified by the work of 
Kurucz, and Castelli (cf. \citet{cas15}).               
But they are an important 
supplement.  Most of the wavelengths in \citet{ko64b} 
are from the laboratory,
while Aller's are from $\gamma$ Geminorum or the Sun.

Subsequent studies of Sirius would use electronic detectors as
well as observations from space.
A recent, detailed
abundance study of Sirius was by \citet{lan11} (henceforth, JDL), 
to which we refer for
an excellent summary of the Sirius system.  Landstreet's work is
based on ESO  Ultraviolet and Visual Echelle Spectrograph 
(UVES) \citep{bag93}, $COPERNICUS$ \citep{rog87}, and 
GHRS \citep{wah93} spectra.  
Detailed synthetic calculations,
described in \citet{fur95}, and published in part as a special report
were described by \citet{kur79}. 
Because of the rotational velocity
of Sirius ($\vsini = 16$ km s$^{-1}$) the effective resolution
of $COPERNICUS$, GHRS, and STIS spectra are comparable.

Sirius was included among the A-type stars studied by \citet{lem89,lem90}.
He determined abundances, using LTE and NLTE methods for C, Fe, Ti,
Si, Ca, Sr, and Ba.  More recently, 
\citet{tak08},
gave abundances of C , O , silicon (Si, $Z$ = 14), titanium (Ti, $Z$ = 22),
Fe , and Ba in 46 A-type stars.
\citet{tak12} subsequently 
determined NLTE abundances  of lithium (Li, $Z$ = 3), sodium (Na, $Z$ = 11), 
and potassium (K, $Z$ = 19) in 24
sharp-lined A-stars, including Sirius.  Earlier, \citet{qiu01} had 
determined abundances of 23 elements in Sirius and Vega. Their work
included Zn, in addition to the Sr, Y, Zr triplet,
Ba and La.
                                                                        
Virtually all of the work cited was restricted to the elements lighter 
than copper (Cu, $Z$ = 29), and the islands of detectability.  But 
efforts to explore the less accessible heavier elements in Sirius have
been made, chiefly based on $COPERNICUS$ but also on $IUE$ spectra.
A study very much in the spirit of the present work is that of
\citet{wah98}, who investigated hot Am stars as a bridge between
HgMn and ordinary Am stars.  They noted explicitly the importance of 
very heavy elements (VHE) as a key to abundance 
patterns.
   
\citet{sad88} used
$COPERNICUS$ spectra to obtain an abundance excess of 
mercury (Hg, Z = 80) in Sirius,
which we confirm.  Subsequently, 
\citet{sad91} 
computed equivalent widths in Sirius for all elements with $Z$ $>$ 40
listed by \citet{kup75} (henceforth, KP), assuming solar abundances.  
He then
reported abundances of molybdenum (Mo, $Z$ = 42), cadmium (Cd, $Z$ = 48),
tungsten (W, $Z$ = 74), and lead (Pb, $Z$ = 82).  We used the same features 
for these elements
as \citet{sad91}, but with the advantage of more recent $gf$-values.
Moreover, numerous atomic or ionic lines are now available that
were not in \citet{cob62}, the resource used by KP for heavier
atoms and ions.  

\citet{yus06} 
reported abundances for 16 elements from Cu to uranium (U, $Z$ = 92): 
Cu, Ga, Y, Zr, Mo, Cd,
tin (Sn, $Z$ = 50), hafnium (Hf, $Z$ = 72), tantalum (Ta, $Z$ = 73), W, 
rhenium (Re, $Z$ = 75), 
osmium (Os, $Z$ = 76), mercury, lead (Pb, $Z$ = 82), 
thorium (Th, $Z$ = 90), 
and U.  Like Sadakane's
work, this was based on $COPERNICUS$ spectra.  
Yushchenko and Gopka show a plot of their
abundances versus $Z$, but their brief contribution does not include the
lines used or designate any points as upper limits.


\section{Observations\label{sec:obs}}
\subsection{STIS spectra\label{sec:stis}}
The characteristics and operational capabilities of {\em HST}\/ Space 
Telescope Imaging Spectrograph (STIS) have been detailed in a number of 
previous publications, especially \citet{woo98} and  
\citet{kim98}.

The STIS observations of Sirius were carried out over a three week period in 
2014 March, as part of a larger project, The Advanced Spectral Library -- Hot 
Stars (ASTRAL)\footnote{see: http://casa.colorado.edu/$\sim$ayres/ASTRAL/}, 
whose aim was to collect high signal-to-noise, full UV coverage 
(1150--3100~\AA) ``atlases'' of 21 representative bright early-type stars 
(types O--A) at the highest spectral resolution feasible with STIS, rivaling 
the quality of ground-based measurements routinely acquired in the optical and 
near-infrared. The specific observing scenario for Sirius was one of 
essentially three set programs for the ASTRAL Project, which were tailored for 
specific types of objects. The scenario -- called 7-Samurai (7-S) -- combined 
seven high- and medium-resolution echelle settings of STIS to achieve the 
highest spectral resolution ($R\equiv\lambda/\Delta\lambda\sim$110,000) in 
specific regions where important stellar, and especially the narrow 
interstellar, features are found; but medium resolution ($R\sim$30,000--45,000)
over other parts of the spectrum where high-S/N might have been unobtainable 
otherwise (in the limited observing time, typically 12 orbits per target for 
the 7-S program).

\begin{deluxetable}{cccccr}
\tabletypesize{\small}
\tablecaption{{\em HST}\/ STIS Observation Log for Sirius\label{tab:obs}}
\tablewidth{0pt}
\tablecolumns{6}
\tablehead{
\colhead{Exp. No.} &  
\colhead{Mode/Setting} &
\colhead{U.T.\ Start} & 
\colhead{$t_{\rm exp}$} & 
\colhead{Aperture} & 
\colhead{QF}\\[3pt]                
\colhead{(1)} &  
\colhead{(2)} &
\colhead{(3)} &  
\colhead{(4)} &
\colhead{(5)} &  
\colhead{(6)}              
}                
\startdata   
\cutinhead{Visit T0: 2014 March 9}
1  & E140H-1271  & 17:59  & $1{\times}1500$  & ND2  &  62 \\
2  & E140M-1425  & 19:08  & $2{\times}1296$  & ND3  &  61 \\
3  & E230M-1978  & 20:44  & $2{\times}1296$  & ND3  & 114 \\
4  & E230H-2713  & 22:19  & $2{\times}1297$  & ND3  & 106 \\
\cutinhead{Visit T2: 2014 March 21}
1  & E140M-1425  & 11:59  & $1{\times}1519$  & ND3  &  46 \\
2  & E230H-2463  & 13:10  & $2{\times}1294$  & ND3  &  86 \\
3  & E230H-2513  & 14:46  & $2{\times}1297$  & ND3  &  93 \\
4  & E230H-2912  & 16:22  & $2{\times}1297$  & ND3  & 110 \\
\cutinhead{Visit T1: 2014 March 30}
1  & E140H-1271  & 06:19  & $1{\times}1500$  & ND2  &  63 \\
2  & E140M-1425  & 07:30  & $2{\times}1296$  & ND3  &  60 \\
3  & E230M-1978  & 09:05  & $2{\times}1296$  & ND3  & 112 \\
4  & E230H-2713  & 10:41  & $2{\times}1297$  & ND3  & 109 \\
\enddata
\tablecomments{Col.(4): $n{\times}t$ format indicates number of 
sub-exposures ($n$) and integration time per sub-exposure ($t$) in 
seconds.  Col.~(5): ND2 is the $0.2^{\prime\prime}{\times}0.05^{\prime\prime}$
ND aperture; ND3 is the $0.3^{\prime\prime}{\times}0.05^{\prime\prime}$ND 
aperture.  These filtered slits have roughly 1\% and 0.1\% transmission, 
respectively, compared with the normal clear echelle apertures.  
Col.~(6): QF is a quality factor, the signal-to-noise per spectral 
resolution element (resel) averaged over the high-sensitivity central 
region of the echellegram.}
\end{deluxetable}

The STIS observing program for Sirius is summarized in Table~\ref{tab:obs}. 
The 12 orbits 
were divided into three 4-orbit ``visits,'' carried out about 10~days apart. 
Visits 0 and 1 were identical, and led off in the first orbit with a standard 
CCD target acquisition using direct imaging (``MIRROR'') through the F25ND5 
heavily-filtered aperture. This was followed by a dispersed-light ``peak-up'' 
(target centering via a raster search) in the $0.3{\times}0.05$ND (ND=3) slit 
again using the CCD, but now with the 4451~\AA\ setting of the G430M 
first-order grating. Filling out orbit 1 was a 1.5 kilosecond (ks) 
high-resolution echelle exposure of the 1200--1350~\AA\ region with E140H-1271 
through the $0.2{\times}0.05$ND (ND=2) slit. Sirius is so bright in the UV 
that these heavily-filtered slits (transmissions of 1\% and 0.1\% for ND2 and 
ND3, respectively) must be used even for the highly dispersed echelle 
spectroscopy. The relative slit positions are known accurately enough that a 
peak-up in ND3 can be transferred to an ND2 spectral observation. In the 
second orbit, a pair of 1.3~ks exposures was taken, covering the broader FUV 
band 1150--1700~\AA\ with a medium-resolution echelle E140M-1425, now through 
the ND3 slit (which also was used for all the subsequent exposures). The third 
orbit was similar, again with a pair of 1.3~ks exposures, but switching to the 
NUV (1700--3100~\AA) with medium-resolution grating setting E230M-1978, whose 
coverage is 1600--2400~\AA. Finally, in orbit 4 a pair of 1.3~ks exposures was 
taken with the high-resolution NUV setting E230H-2713, which covers the range 
2580--2830~\AA, capturing important interstellar absorptions of Fe \II\, 
near 2600~\AA\ and Mg \II\, near 2800~\AA, among the more numerous stellar 
lines.

Visit~2 began with the same CCD imaging acquisition and ND3 slit peak-up as 
visits 0 and 1, but the orbit 1 exposure now was with the medium-resolution 
echelle E140M-1425 and ND3 slit, for 1.5~ks. Orbits 2-4 were devoted to NUV 
high-resolution settings E230H-2463, E230H-2513, and E230H-2912, 
one orbit each and again 
split into equal-length sub-exposures of 1.3~ks, and all through ND3. The 
latter two settings overlap with the blue and red ends of E230H-2713, respectively,
 boosting exposure in the Fe \II\, and {Mg \II\,} interstellar intervals. 
The E230H-2463 setting almost completely overlaps with E230H-2513, and the 
combination was treated effectively as a single setting; however, E230H-2463 
extends just enough to the blue to join the red end of E230M-1978, to ensure 
spectral continuity through the NUV.

The full-orbit exposures (i.e., in orbits 2--4 of a visit) were broken into 
two equal-length sub-exposures in an effort to counteract ``breathing'' 
effects: the telescope focus can change slightly in response to thermal 
cycling and cause the image at the STIS slit focal plane to blur somewhat, 
possibly affecting the throughput, and in extreme cases inducing undesirable 
spectral ``tilts.'' This effect is of particular concern when very narrow 
slits are used, such as the ND2 and ND3 utilized exclusively for the Sirius 
program. To monitor the breathing effect, a quality factor (``QF'' in Table~1) 
was calculated for each exposure by determining the total net counts (from the 
{\sf CALSTIS} pipeline x1d file) in the central roughly two-thirds of the 
echelle pattern, then dividing by the number of independent resolution 
elements (resel) per (truncated) echelle order (1 resel$\sim 2$ spectral bins 
in the $1024{\times}1024$ image format) and by the number of echelle orders. 
The resulting quantity is the average number of net counts per resel for that 
exposure. The square root of that value then is approximately the S/N per 
resel, which was adopted as the quality metric. A large difference between the 
QF values of same-setting sub-exposures in an orbit would signal -- in the 
absence of rapid stellar variability, certainly not anticipated for a star 
like Sirius -- a change in the throughput due to focus variations. In the 
specific example of Sirius, most of the exposure pairs showed some evidence of 
these breathing effects, and it always was the case that the second 
sub-exposure exhibited higher throughput than the first, on average by a 
factor of 1.36${\pm}$0.06. An important side effect of the focus-related 
throughput variations is a slight tilting of the spectrum, because the 
telescope point spread function is wavelength dependent, which becomes most 
conspicuous in the settings that cover the most spectral territory (e.g., 
E230M-1978). These spectral tilting effects must be specially compensated in 
the spectral post-processing.

The STIS echellograms of Sirius were initially processed through the normal 
{\sf CALSTIS} pipeline to yield the so-called x1d file, a tabulation of fluxes 
and photometric errors versus wavelength for each of the up to several dozen 
echelle orders of a given grating setting. This file then was subjected to a 
number of post-processing steps to: (1) correct the wavelengths for small 
errors in the pipeline dispersion relations; (2) de-tilt the spectra, as 
necessary, based on a polynomial correction derived relative to the highest 
throughput sub-exposure of a set (considering all the visits for a given 
target); (3) adjust the echelle sensitivity (``blaze'') function empirically 
based on achieving the best match between fluxes in the overlap zones between 
adjacent orders; and (4) merge the overlapping portions of adjacent echelle 
orders to achieve a coherent 1-D spectral tracing for the specific 
sub-exposure of that setting. Then followed a series of steps to merge the up 
to several independent (visit-level) exposures in a given setting, and finally 
splice these into a full-coverage UV spectrum of the object. An early 
description of these protocols, for the STIS StarCAT\footnote{see: 
http://casa.colorado.edu/$\sim$ayres/StarCAT/} catalog, has been provided by 
Ayres (2010); but the most recent incarnation, including a number of key 
changes and improvements, is described on the ASTRAL site mentioned earlier. A 
fundamental component of the new ASTRAL protocols, as in the earlier StarCAT 
version, is a bootstrapping approach to provide a precise relative, and 
hopefully also accurate absolute, wavelength scale; and similarly for the 
stellar energy distribution. However, with the extensive use of the ND filters 
for Sirius, the absolute flux scale might not be as accurate as for another 
target for which the normal clear apertures, better-calibrated and 
less-affected by breathing, could be used.
\subsection{Ground-based spectra\label{sec:grnd}}

In addition to the STIS observations, we made use of the UVES and 
Kurucz-Furenlid spectra noted in Section~\ref{sec:past}.  Characteristics
of this material are given in the references cited there.  Because the 
Kurucz-Furenlid spectra were photographic, while the UVES reduction was
not optimal, especially for the longer wavelengths, we also used 
NARVAL I spectra downloaded from the
PolarBase~\citep{pet14} site.  This material is described in detail
by~\citep{slv12}.

\section{Identifications, abundances, and upper limits\label{sec:abul}}
\subsection{Wavelengths and preliminary identifications\label{sec:wlid}}

Measured stellar wavelengths, and tentative identifications are 
available for STIS spectra from 
1300.28 to 1999.90 \AA,
\footnote{http://dept.astro.lsa.umich.edu/$\sim$cowley/Sirius/1320newlist.html} 
and 2000.34 to 3044.87 \AA.
\footnote{http://dept.astro.lsa.umich.edu/$\sim$cowley/Sirius/ng20newlst.html}
\newline\noindent Unlike the wavelengths in the papers cited in 
Section~\ref{sec:past}, these
wavelengths were measured directly from the STIS spectrum independently of
individual laboratory or Ritz wavelengths.  They are therefore suitable 
to help decide if an absorption minimum may be attributed primarily
to a single atomic or ionic line. 
Preliminary or tentative identifications
were made with the help of predicted line depths from the VALD3 
\citep{rya15} ``extract stellar'' option,
 which was supplied with $T_{\rm eff}=9900$K and
$\log(g)=4.3$ from JDL.  We enhanced those abundances, often by a factor
of 3 so that lines from exotic species would not be dropped
from the calculation as too weak.   
In subsequent abundance calculations, we used the same $T_{\rm eff}$ and
$\log(g)$, but with JDL's abundances, unenhanced.

A total of 5137 STIS wavelengths between 2000.34 and 3044.87\,\AA\,
were measured.
The measurements were made line-by-line with the help of a visual display
of the normalized spectra.  A cursor was set near minima and a 5-point
least-square parabola fit to the surrounding points.  The adopted
wavelength is the minimum of the parabola.  
For features in
the shoulders of lines, or where no minimum was seen, 
the cursor was set at an (subjective) estimate of where the minimum
might be.  The  
position nearest the cursor was then taken for the wavelength.  
 
We typically consider a deviation of up to 0.02 \AA\, 
of a stellar line from
a laboratory position to be within the error of measurement, or to be
caused by a minor blend.  Larger deviations are taken to indicate
more serious blending.  Quite often, especially in the ultraviolet,
many features will be very close blends--within $\pm$ 0.02\,\AA--often of 
a relatively strong line, e.g. Fe \II, with a line of interest.  An abundance
worker would usually ignore such features, and seek more favorable lines.     
In our cases, we sometimes have no alternative than to 
consider that the line of interest may still make a detectable contribution
to the absorption, and try to determine what it might be.
																																																																				
In many cases, there
was no indication of the presence of the feature sought, and only an upper
limit was assigned.  We will comment on the value of an
upper limit determination below (Section~\ref{sec:theory}).

Few should doubt that atoms of all of the stable elements from Cu and 
Bi, as well as U and Th are present at some level 
in the atmosphere of Sirius.  Some elements may have only lines 
below our current threshold of detectability.
In this work, we may provide at least minimal information on 
such elements by reporting upper limit estimates.
Additionally, we relax the usual criteria
for making line identifications and/or abundance determinations.  
We accept that some atomic or ionic lines may make contributions to the
absorption without having a close, measurable 
wavelength minimum.  This method is not new, and is commonly
used when 
abundances are determined by spectrum synthesis from spectra that
are blended by large values of $\vsini$.  

\subsection{Analysis\label{sec:syn}}

Our abundance analyses have all used spectral synthesis based on a model
atmosphere with temperature, gravity, and abundances adopted by JDL.
The models assumed hydrostatic and radiative equilibrium, and LTE,
and are essentially ATLAS9 models \citep{cak03,kur15}.
In lieu of a
lengthy description and comparison of individual techniques, we list
in Table ~\ref{tab:citat}
representative papers where the different techniques have been 
used.  For the elements studied by two or more authors, many cross
checks were carried out to prevent errors that could arise from
a variety of causes, such as misidentification of 
contributors to blends, misplacement of continuua, or
poor atomic data.  
\begin{deluxetable}{lll}
\tablecaption{Codes and relevant references\label{tab:citat}}
\tablehead{
\colhead{Author}           & 
\colhead{Code(s)}          & 
\colhead{Reference}}           
\startdata
CRC     & ATLAS9/DYNTHL           & \citet{cow10} \\
FC      & ATLAS9/SYNTHE        &\citet{cas15} \\
AFG     & ATLAS9/SYNSPEC49/    &   \\
        & STELLAR              &\citet{hil10}  \\
RM      & ATLAS9/SYNSPEC48/                    \\
        & ROTIN3               &\citet{kil16} \\
GMW     & ATLAS9/SYNTHE           &\citet{wah97} \\
\enddata
\end{deluxetable}

For the elements considered, we have synthesized the region of
one or more of the strongest available lines.  
These were typically chosen with the
help of the NIST online Handbook \citep{san05}.   
We also examined the 
output from VALD3's \citep{rya15} ``extract stellar'' option,
which was supplied with the model and abundances of JDL.
Each of these sources has a distinct advantage.  The VALD3 output supplies
likely blends, along with an estimate of their strengths.  But in the 
STIS region, a number of important lines have antiquated oscillator
strengths.  In a few cases, there were no data in VALD3, 
and the corresponding lines could not be in the originally computed spectra.
Examples are for singly-ionized antimony (Sb, $Z$ = 51) and 
tellurium (Te, $Z$ = 52).    

Typically, our synthesis would proceed with an assumed 
abundance of the element that
was 2 dex (and sometimes more) higher than the solar value.  This was 
useful to show clearly where a line would occur in a region with complex 
blending, because the line of interest would be ``overpredicted,'' that is
stronger than the observed spectrum.  The abundance would then be reduced 
to ``fit'' the observation.
																																																																																				
All calculations have been made in LTE, with default broadening parameters.
Hyperfine structure or isotope shifts were included in calculations for Cu, I, Eu,
Re, Hg, Tl, and Bi.  Typically, the abundances were lower when these effects
were included, by 0.2 to 0.5 dex. 


\subsection{Oscillator strengths and partition functions\label{sec:gfs}}

Oscillator strengths for the 55 individual target elements were
obtained with the help of references from the NIST online
database \citep{kra14}.
Sources for oscillator strengths are listed in Table~\ref{tab:results}.
A glance at the publication dates of these references shows that most 
of this work was unavailable to the researchers mentioned in 
Section~\ref{sec:past}.
In a few cases, we were unable to find satisfactory values, and have used
{\em ad hoc} determinations (this paper), as follows:
\begin{itemize}
\item For Sb \II\, at 1387.56 \AA, the Cowan code \citep{cow81}  was 
used to obtain $\log(gf)= -0.34$.
\item For the line of singly-ionized cesium (Cs, $Z$ = 55), 
Cs \II\, at 4603.79 \AA, we used the value
$\log(gf) = 0.4$,  based on an analogous transition in Kr \I,
at 8113 \AA\, ($5s-5p$).

\end{itemize} 

For background or blending lines,
$gf$-values came from VALD3 or the Kurucz website
\citep{kur13}.  Differences in these sources were occasionally
of significance and were adjusted so that authors used the
same values.

Partition functions occasionally needed reconcilation among
workers.  The main sources for the heavy elements 
were the block data segments from
the SYNTHE code \citep{kur15} or assembled independently
by CRC ( Cowley, cf. \citet{cow03}).

\subsection{Grades of determinations\label{sec:grades}}  
     
The analysis performed with the different codes and line lists
given in Table~\ref{tab:citat} has shown that abundances for
Cu, Zn, elements in the islands of detectability (Sr, Y, Zr, Ba),
Br, Ce, and Pr may be considered known to 0.2 dex or better. 
Abundances for Sr, Y, Zr, and Ba were redetermined
in the present work.  They agreed to 0.05 dex or
better, with
those of JDL (Column L11 of his Table 1).   

The measured wavelengths played an important role in our assessment
of the quality of an abundance determination.
When the wavelengths are close to the laboratory positions, and
some 80\% or more of the absorption is due to the element in question,
abundances are assigned a `g' for good.  However, in some cases,
such as Mo or Nd, the uncertainties range up to 0.3 dex, because of 
difficulties with continuum location, blends, or oscillator
strengths. 

When there is no absorption minimum within 0.02\,\AA, but calculations 
show that roughly half of the absorption is due to the element in 
question, we assign a grade of `f' for fair.  An example
of such a fit is at 1414.40~\AA\, of Ga \II.  In the remainder of cases, we 
assign a grade of poor, `p', or an upper limit `ul'.  We consider the
abundance uncertainties in the `f' to `p' categories to range up to 0.5 dex.

\section{Results\label{sec:results}}
\begin{deluxetable*}{l l l l l r l l}
\tabletypesize{\small}
\tablecaption{Brief tabulation of results\label{tab:results}}
\tablehead{
\colhead{Spect} & 
\colhead{$\lambda$(\AA)$^1$}& 
\colhead{$\lambda^*$(\AA)$^2$} & 
\colhead{Q} & 
\colhead{[El/H]} &
\colhead{Sun$^3$}&
\colhead{$\log(gf)$} &
\colhead{$gf$--Ref.{\Rh}{\Rh}}                              
}
\startdata
Cu \I  &3273.96    &73.98   & g   &   +0.7  & 4.18&$-0.359$ &\citet{liu14} \\
Zn \I  &4722.16    &22.19   & g    &  +1.1  & 4.56&$-0.39$ &\citet{bie80} \\
Ga \II  &1414.40    &14.39   & f   &  +0.7 & 3.02&$+0.25 $ &\citet{shi07} \\  
Ge \II  &1649.19    &49.22   & f   &  +0.5 & 3.63&$-0.28 $ &\citet{fuw05} \\
As \II   &1375.07    &nm$^b$ &ul   &  +1.8& 2.30&$-0.632$ &\citet{bia98} \\ 
Se \I   &1960.89    &60.85   & p   &  +0.6 & 3.34&$-0.434$ &\citet{mor00} \\
Br \I   &1488.45    &88.43   & g   &  +1.3 & 2.54&$-0.58 $ &\citet{mor00} \\
Kr \I   &8776.75    &nm      &ul   &  +3.     & 3.25&$+0.19 $ &\citet{fuh96}\\
Rb \I   &7947.60    &nm      &ul   &  +1.7    & 2.47&$+0.14 $ &\citet{mor00} \\
Sr \II  &4077.71    &77.72   &g    &  +0.7 & 2.83&$+0.15 $ &\citet{fuh96} \\
Y \II   &3600.73    &00.73   &g    &  +0.7 & 2.21&$+0.34 $ &\citet{bie11} \\
Zr \II  &3391.97    &91.97   &g    &  +0.7 & 2.59&$+0.57 $ &\citet{lju06}\\
Nb \II  &3215.59    &nm      &ul   &  +1. & 1.47&$-0.24 $ &\citet{nil10} \\
Mo \II  &2020.31    &20.30   &g    &  +1.0 & 1.88&$+0.022$ &\citet{sil01} \\
Ru \II  &1875.56    &75.54   &p    &  +1.0 & 1.75&$-0.23 $ &\citet{pal09} \\
Rh \II  &1634.72    &nm      &ul   &  +1.5& 0.89&$-0.93 $ &\citet{qui12} \\
Pd \II  &2488.91    &nm      &ul   &  +1.5 & 1.55&$+0.40 $ &\citet{qui96} \\
Ag \II  &2246.41    &nm      &ul   &  +1.1 & 0.96&$+0.42 $ &\citet{kra13} \\
Cd \II  &2265.02    &65.01   &g    &  +0.7 & 1.77&$-0.34 $ &\citet{glo09} \\ 
In \II  &1586.37    &86.33   &ul bl& +1.8 & 0.80&$+0.14 $ &\citet{cur00} \\
Sn \II  &1899.97    &99.90   &p bl & +0.7 & 2.02&$-0.22 $ &\citet{olh10} \\
Sb \II  &1387.56    &nm      &ul   &  +1.3  & 1.01&$-0.337$ &see Sec.~\ref{sec:gfs}\\
Te \II  &1461.68    &61.57   &ul   & +1.  & 2.18&$-0.74 $ &\citet{zha13} \\
I \I    &1457.98    &57.98   &g    &  +2.0  & 1.55&$+0.20 $ &\citet{cha10} \\
Xe \I   &1469.61    &nm      &p    & +1.1 & 2.24&$-0.58 $ &\citet{mor00} \\
Cs \II  &4603.79    &nm      &ul   & +3.8 & 1.08&$+0.40 $ &see Sec.~\ref{sec:gfs} \\
Ba \II   &4554.03    &54.05   &g   &  +1.4  & 2.25&$+0.14 $ &\citet{klo02} \\
La \II  &4042.91    &nm      &f    &  +1.5  & 1.11&$+0.33 $ &\citet{zhi99} \\
Ce \II  &4460.21    &nm      &g    &  +1.5  & 1.58&$+0.28 $ &\citet{law09} \\                               
Pr \II  &4222.93    &nm      &g    &  +2.2  & 0.72&$+0.27 $ &\citet{bie03} \\
Nd \III &5294.10    &nm      &g    &  +1.4  & 1.42&$-0.69 $ &\citet{rya06} \\
Sm \II  &3568.27    &nm      &f    & +2.       & 0.95&$+0.29 $ &\citet{law06} \\
Eu \II  &4205.05    &nm      &f    &  +1.3  & 0.52&$+0.21 $ &\citet{la01a} \\
Gd \II  &4251.73    &51.76   &p    &  +1.6  & 1.08&$-0.22 $ &\citet{den06} \\
Tb \II  &3509.15    &nm      &p    & +2.  & 0.31&$+0.70 $ &\citet{la01b} \\
Dy \II  &3531.70    &31.61   &p    & +1.7 & 1.10&$+0.77 $ &\citet{wic00} \\
Ho \II  &3456.02    &nm      &ul   & +1.8 & 0.48&$+0.76 $ &\citet{law04} \\
Er \II  &3499.10    &nm      &f    & +1.8 & 0.93&$+0.29 $ &\citet{law08} \\
Tm \II  &3462.20    &nm      &ul   & +2.1 & 0.11&$+0.03 $ &\citet{wic97} \\
Yb \II  &3289.37    &89.39   &ul   & +1.8 & 0.85&$-0.05 $ &\citet{bie98} \\
Lu \II  &2615.42    &15.42   &f    & +1.6 & 0.10&$+0.14 $ &\citet{qui99} \\
Hf \II  &2647.29    &nm      &ul   & +1.8 & 0.85&$+0.46 $ &\citet{bou15} \\
Ta \II  &2400.13    &00.03   &ul   & +1.7 &$-$0.12&$+1.65$ &\citet{cob62} \\
W \II   &2029.99    &29.96   &p    &  +1.5  & 0.83&$+0.18 $ &\citet{nib08} \\
Re \II  &2275.25    &nm      &ul   &  +1.7  & 0.26&$-0.385$ &\citet{ort13} \\
Os \II  &2194.39    &94.41   &f    &  +1.0  & 1.40&$-0.22 $ &\citet{qui06} \\
Ir \II  &2126.81    &26.85   &ul   & +1.0 & 1.42&$+0.11 $ &\citet{xup07} \\
Pt \II  &1883.06    &83.04   &f    &  +1.6  & 1.62&$-0.12 $ &\citet{qui08} \\
Au \II  &2082.07    &82.05   &f    & +1.5 & 0.91&$-0.09 $ &\citet{fiv06} \\
Hg \II  &1942.27    &42.31   &g    &  +1.2  & 1.17&$-0.418$ &\citet{pro99} \\
Tl \II  &1321.64    &21.68   &f    &  +0.9  & 0.90&$+0.12 $ &\citet{cur00} \\
Pb \II  &2203.53    &03.49   &f    &  +1.4  & 1.92&$-0.143$ &\citet{mor00} \\
Bi \II  &1791.84    &nm      &ul   &  +1.6  & 0.65&$-0.51 $ &\citet{pal01} \\
Th \II  &4019.13    &nm      &ul   & +1.5 & 0.03&$-0.23 $ &\citet{nil02} \\
U  \II &4241.66     &nm      &ul   &  +3.   & $-$0.54&$-0.10$ &\citet{niv02} \\
\enddata
\tablenotetext{1}{Wavelength, vacuum $\lambda < 2000$\AA, air above}
\tablenotetext{2}{First 2 digits of measured stellar wavelengths omitted}
\tablenotetext{3}{$\log(H) = 12.00$}
\tablenotetext{b}{No measurement; no line close}
\end{deluxetable*}		
																																																														
Table~\ref{tab:results} is a concise summary of 
our results for the 55 elements.  The elemental abundances are given
as logarithmic ratios of elements-to-hydrogen, on the astronomical
scale where $\log(H) = 12$. All entries in the column labeled [El/H]
were newly determined in the present work. 

The solar abundances are from recent updates of 
\citet{scb15}, and \citet{gre15}.  For elements not in these
updates, values were taken from \citet{asp09}, Table 1.  
Meteoritic abundances assigned by \citet{asp09} were
used when photospheric values were not available. 
Of the 14 abundance results considered good (g), most are for species 
determined by other workers, notably JDL, or \citet{sad91}.  Twelve of the abundances
are considered fair (f), while eight are marked poor (p).  For 21 elements, we
give only upper limits (ul).

Space does not permit a discussion of 55 individual
elements, but for a few elements we discuss
specific cases, and examples are given for the 
various categories of quality.  Detailed information for individual elements
may be found on a website dedicated to the Sirius analysis:
http://dept.astro.lsa.umich.edu/$\sim$cowley/Sirius/.
Table~\ref{tab:linx} lists additional lines that were examined to 
support the abundances listed in Table~\ref{tab:results}.

In the sample plots of the following sections, the lines of the target 
elements typically have better fits than neighboring features. This happens 
primarily because (1) we select the regions best suited to the target element, 
and (2) for the latter, we adjust the abundance for an optimum fit. Random 
wavelengths especially in the ultraviolet, often have poor oscillator 
strengths. Additionally, these regions are replete with line absorption whose 
origin is not known.

\subsection{Sample: Bromine (Br, Z = 35)\label{sec:Br}}  

\begin{figure}
\epsscale{1.2}
\plotone{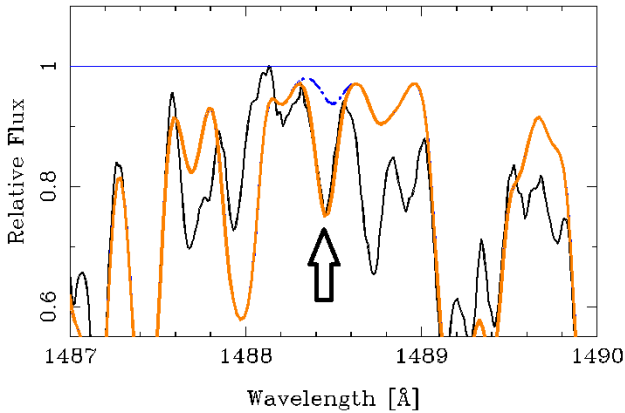}
\caption{Synthesis of the region near Br \I\, at 1488.45 \AA.
The broken (blue online) line was computed with a solar Br abundance.  The 
gray (orange online) spectrum assumed an enhancement of 1.3 dex above 
solar.  This determination is rated: `g' (good).\label{fig:Br1488}}
\end{figure}

Because Br has a first ionization energy of 11.8 eV,
it is reasonable to expect the resonance lines of Br \I\, would appear.  Lines
at 1488.45 and 1540.65\,\AA\, arise from the ground state.  

Synthesis of the line at 1488 \AA\, is shown in Fig.~\ref{fig:Br1488}.
The broken (blue online) line was computed assuming a solar bromine abundance, 
while the gray (orange online) line was made 
assuming a 1.3 dex enhancement.  The measured stellar 
wavelength was 1488.434\,\AA,  
and the synthesized fit is excellent.  The oscillator strength is from
\citet{mor00}.  Of two possible lines
that might confirm the identification, one, at 1540.65\,\AA\, is at a region with
missing absorption.  However, there is no measured minimum at this wavelength,
and additional contributors are needed to explain the absorption.  Another
line, at 1574.84\,\AA\, is  significantly weaker than the other two.
Moreover, it is closely coincident with strong lines of Si \I, measured 
at 1574.87 \AA.
  
We conclude that the bromine abundance in Sirius is 1.3 dex above solar.
The determination is graded `g'.

\subsection{Sample: Gallium (Ga, $Z$ = 31)\label{sec:Ga}}

\begin{figure}
\epsscale{1.2}
\plotone{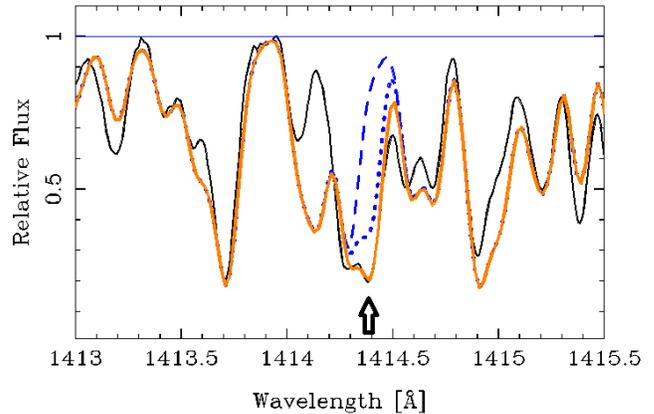}
\caption{Synthesis of the region near Ga \II\, at 1414.40 \AA.
The dotted (blue online) line was computed with a solar Ga abundance.  The
dashed line was computed assuming no Ga at all. 
The gray (orange online) spectrum assumed an enhancement of 0.7 dex above 
solar.  This determination is rated: `f' (fair).\label{fig:Ga1414}}
\end{figure}

Fig.~\ref{fig:Ga1414} shows the Ga \II\, line at 1414.40 \AA\, closely blended
primarily with Fe \II\ and Ni \II\, on the short wavelength side.  The minimum, indicated
by the arrow, was measured at 1414.39 \AA, in excellent agreement with the
NIST wavelength.  The solid gray (orange online) line was calculated with a 
Ga excess of 0.7 dex, while the dotted line shows the result for a solar
abundance of gallium.  For the dashed blue line, the Ga abundance was set to zero.  
Clearly a substantial portion of the absorption is due to Ga.  We grade this 
determination `f'.

\subsection{Sample: Ruthenium (Ru, $Z$ = 44)\label{sec:Ru}}    

\begin{figure}
\epsscale{1.2}
\plotone{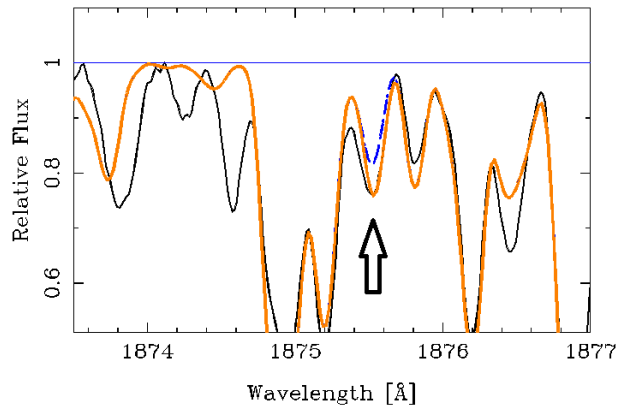}
\caption{Synthesis of the region near Ru \II\, at 1875.56 \AA.
The Ru abundance is enhanced by 1.3 dex for the computed 
gray (orange online) spectrum.  Zero Ru was assumed for the broken 
(blue online) spectrum.
Syntheses near Ru \II\, lines at 1939.04, 1939.51, 1966.07, and 
1966.73 \AA\, give an average of 1.0 as the abundance excess. We rate the
overall determination as `p' (poor). \label{fig:Ru1875}}
\end{figure}

In Fig.~\ref{fig:Ru1875}, the arrow marks the position of the Ru \II\, line, and the
dashed (blue online) spectrum was calculated 
assuming no ruthenium at all.  A solar abundance 
of ruthenium hardly changes the calculation from the dashed line.
That is because of a relatively strong Fe \II\, line at 1875.53~\AA.
VALD3 gives the source of the log($gf$) for Fe \II\, as K13 \citep{kur13}.
We find no other source for that line.  If we use the default log($gf$)'s
for the Fe \II\, and Ru \II, the ruthenium abundance that fits gray (orange online) 
has an excess of 1.2 dex.

We examined the region near two other Ru \II\, lines, 
at 1939.04 and 1939.51 \AA.  These led us to lower our estimate of
the Ru abundance to +1.0 dex above the solar value.  The determination
is graded `p'.  

\subsection{Sample: Rhenium (Re, $Z$ = 75)\label{sec:Re}}  

\begin{figure}
\epsscale{1.2}
\plotone{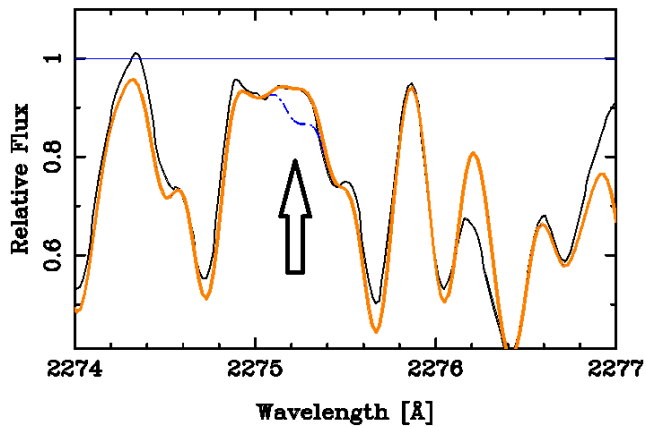}
\caption{Synthesis of the region near Re \II\, at 2275.25 \AA.           
The gray line (orange online) was made assuming an Re excess of 1.7
dex.  The broken (blue online) was made assuming a 2.7 dex excess.
The calculation includes full hyperfine structure.
Only an `ul' (upper limit) was assigned.\label{fig:Re2275}}
\end{figure}

We found only an upper limit to the Re abundance.  Figure~\ref{fig:Re2275}
is illustrative of upper limits.  The gray (orange online) line fits the 
observation roughly with an assumed Re excess of 1.7 dex.  
Since there is
no observed minimum, we report 1.7 dex as the upper limit.  The broken blue
line was calculated assuming a 2.7 dex excess.  Both calculations used the
full hyperfine structure of the 2275 \AA\ line as given by \citet{wah97}.

\vspace{1.0cm}
\subsection{Sample: Iodine (I, $Z$ = 53)\label{sec:I}}    

We are unaware of a credible identification of
iodine in the spectrum of another star, and therefore highlight
this result.  This resonance transition is the analogue of 
the line for Br \I.

\begin{figure}             
\epsscale{1.2}
\plotone{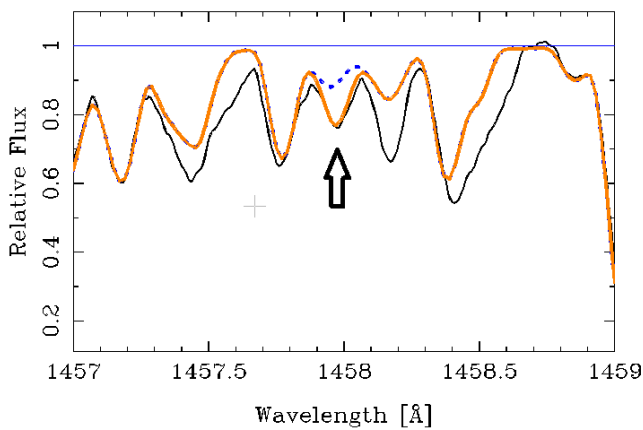}
\caption{Synthesis of the region near I \I\, at 1457.98 \AA.
The I abundance is enhanced by 2.0 dex for the gray (orange online)
calculation.  The broken spectrum (blue online) was made with
solar abundance.  The full 12-component hyperfine structure
was included from \citet{luc75}. 
This determination is rated `g' (good).
\label{fig:II1458}}
\end{figure}

Figure~\ref{fig:II1458} shows I \I\, at 1457.98 \AA.
The stellar and laboratory wavelength measurements agree to
0.01\AA.  However, possible confirmation from  expected lines  
at 1702.07, 1782.74 or 1830.38 \AA\, is unavailable because of blending.
The fit shown in gray (orange online) was made with an excess of 2.0 dex.
The abundance determination is 
graded `g' based on our adopted criteria.  The large abundance excess
makes iodine an ostensible odd-Z anomaly.  However, it is also
possible that the oscillator strengths of closely blending 
lines of Fe \II, Mn \II, and Cr \II\, are underestimates.   This
is likely because of the complex nature of the transitions
themselves.

\section{The abundance pattern of Sirius and its meaning\label{sec:mean}}


\begin{figure}
\epsscale{1.2}
\plotone{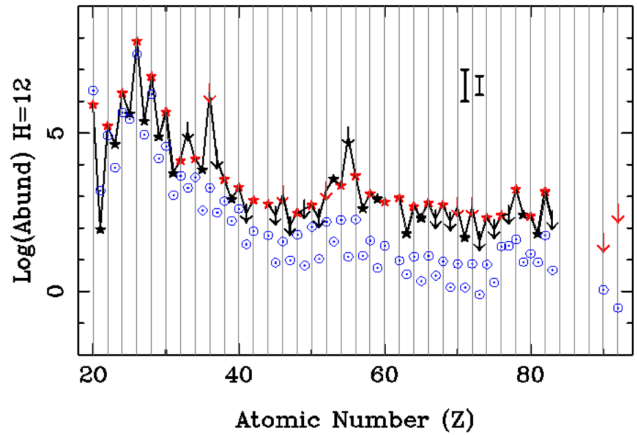}
\caption{Abundances (stars) and upper limits (arrows) for $\alpha$ CMa.  
Even-$Z$ elements are in red.  Solar
abundances are also shown(Sun symbols, blue online).
Error bars are shown for quality good ($\pm 0.3$) and fair to poor
($\pm 0.5$ dex). These uncertainties do not apply to the upper limits.
\label{fig:last}}
\end{figure}

Figure ~\ref{fig:last} shows the abundances and upper limits from Ca ($Z$ = 20)
through U, along with the solar abundances.  The latter
fall typically 1 to 2 dex below those for Sirius.  There are
no indications of the 4 or more dex excesses found in the HgMn stars,
apart, possibly, for Cs ($Z$ = 55). 

The high upper limits at $Z$ = 36 and 55  are
due to the atomic properties of Krypton (Kr $Z$ = 36) and Cs.  
The resonance lines of both Kr \I, Kr \II, and Cs \II\, 
fall in the far ultraviolet, short of our
wavelength coverage. The Cs \I\, lines are accessible,  
but the ionization energy of neutral cesium is so low
(3.89 eV) that observations of the neutral species are precluded
in an A star.  Thus, we cannot assume high upper limits to
be likely  
abundance peaks, though we strictly cannot exclude
that possibility.  The point for I ($Z$ = 52) could indicate
an odd-Z anomaly (cf. Section~\ref{sec:I}).

We see that a number of the exotic elements, that are often 
overabundant by 3 or more
dex in HgMn stars, Ga, Au, Pt, Hg, fall smoothly among their neighbors.
Pt and Pb have abundances 1.6 and 1.4 dex above their 
solar values.  

For typical HgMn stars the Mn abundance increases
with increasing temperature, while the platinum abundance
has been found to be significantly anticorrelated
with temperature \citep{kcs93}.  If, as has been suggested
(e.g. \citet{ade03}), the HgMn and Am stars are
related, we might expect Pt to be unusually high in Sirius.   Its
$T_{\rm eff} = 9900$K would place it 
at the cool end of the HgMn stars. 
For example, the Pt 
excess in the cool ($T_{\rm eff} = 10750$K) HgMn star HR 7775
is 4.6 dex \citep{wah00}.  We find only 1.6 dex for Sirius,
a typical value for most of the heavy elements.

Unfortunately, our upper limits contain no information on possible 
underabundances.  Perhaps the most notorious underabundance of the 
HgMn stars is that of Zn, which is some 4 orders of magnitude below
the solar value in $\chi$ Lupi \citep{lec99}, and
only upper limits were reported for a number of HgMn stars by 
\citet{kcs93}.  While a few HgMn and related stars have Zn excesses
of 1 to 2 dex, as we find for Sirius, Zn in Sirius is not typical
of most HgMn stars.

The STIS results are consistent with the prevailing interpretation of
Sirius as a ``hot Am'' star, with underabundances of certain lighter
elements, and mild excesses of heavier ones.

\section{Comparison of observations to theoretical predictions}
\label{sec:theory}

For nearly half a century, the (only) theory considered likely 
for abundance anomalies
in all varieties of CP stars has been endogenous 
chemical separation--photospheric
or sub-photospheric diffusion, or stellar winds \citep{ale05}.
The models upon which theoretical predictions are based have
evolved significantly since the early 1970's.  In the beginning,
a decisive factor was the ratio of upward (radiation, $g_R$) 
to downward (gravitational and thermal) 
forces in the outer stellar envelopes.
In a figure that might have been relevant for the present paper,
\citet{mic76} gave this ratio for elements with atomic numbers
$Z$ through ca. 65, for stars of 1.2 through 3.3 $M_\odot$.  The plots
appeared to explain the underabundances of Ca and Sc in Am stars,
but showed lack of support for Fe and Y, and Sr conflicting with
the observation that these elements are typically overabundant
in Am stars.

More recent calculations by \citet{mic11} (henceforth MRV) were 
made specifically for Sirius.  Unfortunately, they 
extend only from He through Ni ($Z$ = 28).  Unlike the calculations of
the 1970's, these predictions are based on chemical separations
throughout the bulk of the star and as a function of age. Results are
presented for two basic models, one with a turbulent envelope below
a mixed outer 
mass of ca. $10^{-6}M_\odot$.
A second model did not have turbulence in the envelope, but assumed 
a mass loss near 
$10^{-13}M_\odot$ year$^{-1}$.  The predicted abundance patterns from
the models are remarkably similar (see Figure 4 of MRV).
The authors state that ``Among the 17 abundances determined observationally,
up to 15 can be predicted to within 2$\sigma$, and 3 of 4 determined 
upper limits are satisfied.''  This is a ``glass half full'' conclusion.

A less favorable evaluation could come from a Chi-squared comparison.
This would show that the deviations of JDL's abundances, as cited
earlier, from MRV's
predictions cannot be accounted for in terms of the abundance 
errors ($\le 0.2$ dex).
The Chi-squared sum would be dominated by the upper limit found
for the element B ($Z$ = 5).
That value alone would be sufficient to allow one to reject a null 
hypothesis that the deviations from theory are
due only to observational errors.    
We find independently, an upper limit for B that is 
1.45 dex {\em below} the predicted value.
It is entirely valid to use this as an abundance with the understanding that
a definite value would only increase the Chi-squared, making the
null hypothesis even less acceptable.

MRV note the difficulty with B, and also one with Na.  They suggest that
mass transfer from the secondary could be relevant.  Similar remarks
were made by \citet{ric00} and JDL, who also remarks that the separation
of the two stars has ``led people to assume that no important interaction
occurred.''

\section{Summary\label{sec:summary}}

In this work we have determined abundances for 34 
elements from copper through uranium.  Additionally, 21 upper limits were assigned.
We are able to rule out the possibility that any elements (apart from Cs) have
excesses of 4 dex or more. Abundance excesses of heavy elements fall generally
between 1 and 2 dex.  We find no clear violations of the odd-even
effect as is found in some HgMn stars, for example, at yttrium.  
The most modern theoretical calculations (of MRV)
relevant to Sirius do not cover elements examined here.

\acknowledgments

We gratefully acknowledge the help and advice of R. L. Kurucz, who made
available digital and hard copy
versions of his Sirius spectra and their syntheses.  S. Bagnulo kindly 
supplied usable portions of the UVES spectra of sirius. 
We acknowledge use of NARVAL spectra
from the PolarBase archive \citep{pet14}.  
We also acknowledge use of the Mikulski
Archive for the Space Telescope (MAST) for access to the $COPERNICUS$ and
HST/GHRS material used in this work.
We are thankful for the use of the
NIST online Atomic
Spectroscopy Data Center \citep{kra14}.  G. Nave of NIST has given advice and
clarification on questions of atomic structure and spectra.
This work has also made use of the VALD database, operated at Uppsala University, 
the Institute of Astronomy RAS in Moscow, and the University of Vienna
\citep{rya15}.

\appendix

\section{Appendix: Lines examined\label{sec:linx}}

\begin{deluxetable*}{l l l l| l l l l}
\tabletypesize{\normalsize}
\tablecaption{Additional lines used for abundances\label{tab:linx}}
\tablehead{
\colhead{Element} & 
\colhead{Spectrum} &
\colhead{$\lambda$(\AA)}& 
\colhead{$\lambda$(\AA)} &
\colhead{Element} &
\colhead{Spectrum} &
\colhead{$\lambda$(\AA)}& 
\colhead{$\lambda$(\AA)}  
}
\startdata
Cu & \I & 3247.54 &       &   Zn & \I & 4680.14  & 4810.53 \\
   &    &         &       &   Zn & \II& 2025.48  & \\  
Ga &\III& 1495.04 &       &   Ge & \I & 2041.71  &2068.66  \\
   & \I & 2094.26 &       &   As & \I & 1890.42  & 2860.44 \\
Br & \I & 1540.65 &1574.84&   Rb & \I & 7800.27  &  \\
   &    &         &       &   Rb & \II& 4244.40  & \\
Sr & \II& 4215.52 &       &   Y  &\II & 3242.27  & 3611.04 \\
   &    &         &       &   Y  &\II & 3710.30  &  \\
Zr & \II&3438.23  &3479.39&   Zr &\II & 3496.21   &3551.95     \\
Nb & \II& 2033.01 &2697.03&      &    &          &       \\
Nb & \II& 2109.43 &       &      &    &          &  \\
Ru & \II& 1939.04 &1939.51&      &    &          & \\
   & \II& 1966.07 &1966.73&   Rh &\II & 2420.97  &   \\
Cd &\II & 2144.41 &       &   Sn &\II & 1400.47  &   \\
Sb &\I  & 2068.33 &       &   Te & \I & 2002.03  &    \\
   &    &         &       &   Te &\II & 4654.37  & 5649.26 \\
   &    &         &       &   Te &\II &  5708.12 &     \\
I  &\I  & 1702.07 &1782.74&   Xe &\II & 5419.15  & \\
Ba &\II & 4934.08 &       &   Ce &\II & 4562.36  &    \\
Pr &\II & 4062.81 &       &      &    &          &   \\
Pr &\III& 5299.99 &       &   Nd &\II & 4303.57 & \\
Sm &\II & 3592.60 &       &   Eu &\II & 4129.70   &  \\
   &    &         &       &   Eu &\III& 2375.46   &2444.38 \\
   &    &         &       &   Eu &\III& 2513.76  \\
Gd &\II & 3545.80 &3549.36&  \\
Gd &\III& 2223.95 &3545.80&   Dy &\II & 3436.02 & 3534.96 \\
Ho &\II & 3398.95 &3796.75&  \\
Ho &\II & 3810.74 &       &   Er &\II & 3264.78 & 3372.71 \\
   &    &         &       &   Er &\II & 3616.56   \\
Tm &\II & 3131.26 &3701.36&                            \\
Tm &\II & 3761.33 &       &  Yb  &\III& 2516.81   \\
Lu &\III& 2236.18 &2603.35&        \\
Lu &\III& 2911.39 &       &  Hf  &\II & 3399.79  & 3561.66  \\
Ta &\II & 2146.87 &       &  W   &\II & 1951.05  \\
Re &\II & 2608.50 &       &  Os  &\II & 2067.21  & 2070.67  \\
   &    &         &       &  Os  &\II & 2282.26  \\
Ir &\II & 2169.42 &2579.48&  Pt  &\II & 1777.09  & 2144.25 \\
Au &\II & 1362.33 &       &  Hg  &\II & 1649.94  &   \\
Tl &\II & 1908.62 &       &  Pb  &\II & 1433.91  &1822.05  \\
Bi &\II & 1436.81 &       &  Th  &\III& 1307.44  &1356.92 \\
U  &\II & 1358.57 &4241.66&  U   &\III & 3565.73  \\
\enddata
\tablenotetext{1}{Laboratory wavelength, vacuum $\lambda < 2000$\AA, air above}
\tablenotetext{2}{First 2 digits of measured stellar wavelengths omitted}
\tablenotetext{3}{$\log(H) = 12.00$}
\tablenotetext{a}{No measurement; no line close}
\end{deluxetable*}

Most of the abundance results reported in Table~\ref{tab:results} depend
primarily on a single feature.  But typically several features were
examined for consistency with the adopted values.  
In Table~\ref{tab:linx}, we list the wavelengths
of the lines examined for each element that are not in Table~\ref{tab:results}.  





\begin{thebibliography}{}
\bibitem[Adelman, et al.(2003)]{ade03} Adelman, S. J., Adelman, A. S.
\& Pintado, O. I. 2003, A\&A, 397, 267
\bibitem[Adelman, et al.(2006)]{ade06} Adelman, S. J., Caliskan, H., 
Gulliver, A. F. \& Teker, A. 2006, A\&A, 447, 685
\bibitem[Alecian, Richard \& Vauclair(2005)]{ale05}  Alecian, G., Richard, O.,
and Vauclair, S. (eds) 2005, {\em Element Stratification in Stars: 40 Years
of Atomic Diffusion}, EAS Pub. Ser. 17
\bibitem[Aller(1942)]{all42} Aller, L. H. 1942, \apj, 96, 321
\bibitem[Asplund, et al. (2009)]{asp09} Asplund, M., Grevesse, N., Sauval, J.
2009, ARAA, 47, 481
\bibitem[Ayres(2010)]{ayr10} Ayres, T. 2010, ApJS, 187, 149
\bibitem[Bagnulo, et al. (2003)]{bag93} Bagnulo, S., et al. 1993, ESO
Messenger, 114, 10
\bibitem[Bi\'{e}mont \& Godefroid(1980)]{bie80} Bi\'{e}mont, \'{E}. \& Godefroid,
M. 1980, A\&A, 84, 361 
\bibitem[Bi\'{e}mont, et al. (1998a)]{bia98}  Bi\'{e}mont, \'{E}., Morton, D. C.
\& Quinet, P. 1998, MNRAS, 297, 713  
\bibitem[Bi\'{e}mont, et al. (1998b)]{bie98} Bi\'{e}mont, \'{E}., Dutrieux, J.-F.,
Martin, I. \& Quinet, P. 1998, J. Phys. B, 31, 3321
\bibitem[Bi\'{e}mont, et al. (2003)]{bie03} Bi\'{e}mont, \'{E}., Lef\`{e}bvre, P.-H.,
Quinet, P., et al. 2003, Eur. Phys. J. D, 27, 33
\bibitem[Bi\'{e}mont, et al. (2011)]{bie11} Bi\'{e}mont, \'{E}., Blagoev, K., 
Engstr\"{o}m, L., et al. 2011, MNRAS, 414, 3350
\bibitem[Bouazza, et al. (2015)]{bou15} Bouazza, S., Quinet, P. \& Palmeri, P.
2015, JQSRT, 163, 39
\bibitem[Castelli \& Kurucz (2003)]{cak03}Castelli, F. \& Kurucz, R. L. 2003, in 
Modelling of Stellar Atmospheres, IAU Symp. 210, ed. N. E. Piskunov, 
W. W. Weiss \& D. F. Gray.
\bibitem[Castelli(2015)]{cas15} Castelli, F. 2015,
   http://wwwuser.oats.inaf.it/castelli/hd175640/hd176540.html
\bibitem[Chang, et al (2010)]{cha10} Chang, Z., Li, J. \& Dong, C. 2010, J. Phys.
Chem. A, 114, 13388 
\bibitem[Conti(1965)]{con65} Conti, P. S. 1965, ApJ, 142, 1594
\bibitem[Corliss \& Bozman(1962)]{cob62} Corliss, C. H. \& Bozman, W. R. 1962,
Experimental Transition Probabilities for Spectral Lines of Seventy Elements, NBS
Monog. 53 (CB)
\bibitem[Cowan (1981)]{cow81} Cowan, R. D. 1981, The Theory of Atomic Structure
and Spectra (Berkeley: Univ. Calif. Press)
\bibitem[Cowley, et al. (2003)]{cow03} Cowley, C. R., Adelman, S. J. \& Bord,
D. J. 2003, in Modelling of Stellar Atmospheres, IAU Symp. 210, ed. N. Piskunov,
W. W. Weiss \& D. F. Gray, p. 261
\bibitem[Cowley, et al. (2010)]{cow10} Cowley, C. R., Hubrig, S., 
Gonz\'{a}lez, J. F. \& Savanov, I. 2010, A\&A, 523, 65
\bibitem[Curtis (2000)]{cur00} Curtis, L. J. 2000, Phys. Scr., 62, 31
\bibitem[Den Hartog, et al. (2006)]{den06} Den Hartog, E. A., Lawler, J. E., 
Sneden, C. \& Cowan, J. J. 2006, ApJS, 167, 292
\bibitem[Fivet, et al. (2006)]{fiv06} Fivet, V., Quinet, P., Bi\'{e}mont, \'{E} \&
Xu, H. L. 2006, J. Phys. B, 39, 3587    
\bibitem[Folsom, et al. (2012)]{fol12} Folsom, C. P., Bagnulo, S., Wade, G. A., 
et al. 2012, MNRAS, 422, 2072
\bibitem[Fuhr \& Wiese(1996)]{fuh96} Fuhr, J. R. \& Wiese, W. L. 1996, in CRC Handbook
of Chemistry and Physics, 77th ed. D. R. Lide, ed. (Boca Raton, FL, CRC Press Inc.)
\bibitem[Fuhr \& Wiese(2005)]{fuw05} Fuhr, J. R. \& Wiese, W. L. 2005, in CRC Handbook
of Chemistry and Physics, 86th ed. D. R. Lide, ed. (Boca Raton, FL, CRC Press Inc.)    
\bibitem[Furenlid, et al. (1995)]{fur95} Furenlid, I., Westin, T.
\& Kurucz, R. L. 1995, in Laboratory and Astronomical High Resolution
Spectra, ASP Conf. Ser. 81 (ed. A. J. Sauval, R. Blomme \& N. Grevesse), p. 615
\bibitem[Glowacki \& Migdalek(2009)]{glo09} Glowacki, L. \& Migdalek, J. 2009, Phys.
Rev. A, 80, 042505    
\bibitem[Grevesse, et al. (2015)]{gre15} Grevesse, N., Scott, P., Asplund, M., et al.
2015, A\&A, 573, 27G
\bibitem[Hill \& Landstreet (1993)]{hil03} Hill, G. \& Landstreet, J. D. 1993,
A\&A, 276, 142 
\bibitem[Hill, et al. (2010)]{hil10} Hill, G., Gulliver, A. F. \& Adelman, S. J.
2010, ApJ, 712, 250
\bibitem[Kili\c{c}o\u{g}lu, et al. (2016)]{kil16} Kili\c{c}o\u{g}lu, T., Monier, R.,
Richer, J., et al. 2016, AJ, 151, 49
\bibitem[Kimble et al.(1998)]{kim98} Kimble, R.~A., 
Woodgate, B.~E., Bowers, C.~W., et al.\ 1998, \apjl, 492, L83
\bibitem[Klose, et al. (2002)]{klo02} Klose, J. Z., Fuhr, J. R. \& Wiese, W. L. 2002,
J. Phys. Chem. Ref. Data, 31, 217 
\bibitem[Kohl(1964a)]{koh64} Kohl, K. 1964a, Zs.f.Ap., 60, 115
\bibitem[Kohl(1964b)]{ko64b} Kohl, K. 1964b, Das Spectrum des Sirius (Thesis,
University of Kiel)
\bibitem[Kramida (2013)]{kra13} Kramida, A. 2013, J. Res. Natl. Inst. Stand. Tech.,
118, 168
\bibitem[Kramida, et al. (2014)]{kra14} Kramida, A., et al. 2014, NIST Atomic Spectra
Database (version 5.2), 
[Online]. Available:  http://physics.nist.gov/asd
\bibitem[Kurucz \& Peytremann (1975)]{kup75} Kurucz, R. L. \& Peytreman, E. 1975,
Smithsonian Ap. Obs. Spec. Rept. 362 (KP)   
\bibitem[Kurucz \& Furenlid(1979)]{kur79} Kurucz, R. L. \& Furenlid, I. 1979, Sample
Spectral Atlas for Sirius, SAO Spec. Rept. 387   
\bibitem[Kurucz (2013)]{kur13} Kurucz, R. L. 2013, 
http://kurucz.harvard.edu/linelists/gfnew/gfall18feb16.dat
\bibitem[Kurucz (2015)]{kur15} Kurucz, R. L. 2015
http://kurucz.harvard.edu/programs/synthe/ 
\bibitem[Landstreet(2011)]{lan11} Landstreet, J. D. 2011,
A\&A, 528, 132 (JDL)
\bibitem[Lawler, et al. (2001a)]{la01a} Lawler, J. E., Wickliffe, M. E., Den Hartoog,
E. A. \& Sneden, C. 2001a, ApJ, 563, 1075
\bibitem[Lawler, J. E. (2001b)]{la01b} Lawler, J. E., Wickliffe, M. E., Cowley, C. R.
\& Sneden, C. 2001b, ApJS, 137, 341
\bibitem[Lawler, et al.(2004)]{law04} Lawler, J. E., Sneden, C. \& Cowan, J. J. 2004,
ApJ, 604, 850
\bibitem[Lawler, et al. (2008a)]{law06} Lawler, J. E., Den Hartoog, E. A., Sneden, C.
\& Cowan, J. J. 2006, ApJS, 162,277
\bibitem[Lawler, et al. (2008)]{law08} Lawler, J. E., Sneden, C., Cowan, J. J.,
et al. 2008b, ApJS, 178, 71
\bibitem[Lawler, et al. (2009)]{law09} Lawler, J. E., Sneden, C., Cowan, J. J., et al.
2009, ApJS, 182, 51
\bibitem[Leckrone, et al.(1999)]{lec99} Leckrone, D. S., Proffitt, C. R., 
Wahlgren, G. M., et al. 1999, AJ, 117, 1454
\bibitem[Liu, et al (2011)]{liu11} Liu, Y. P., Gao, C., Zeng, et al.
2011, A\&A, 536, 51 
\bibitem[Liu, et al.(2014)]{liu14} Liu, Y. P., Gao, C., Zeng, et al.
\& Shi, J. R. 2014, ApJS, 211, 30
\bibitem[Lemke (1989)]{lem89} Lemke, M. 1989, A\&A, 225, 125
\bibitem[Lemke (1990)]{lem90} Lemke, M. 1990, A\&A, 240, 331
\bibitem[Ljung, et al.(2006)]{lju06} Ljung, G., Nielsson, H., Asplund, M., et al. 2006,
A\&A, 456, 1181
\bibitem[Luc-Koenig, et al. (1975)]{luc75} Luc-Koenig, E., Morillon, C. \& Verg\`{e}s, J.
1975, Phys. Scr., 12, 199
\bibitem[Michaud (1970)]{mic70} Michaud, G. 1970, ApJ, 160, 641
\bibitem[Michaud, et al. (1976)]{mic76} Michaud, G. Charland, Y., Vauclair, S.
\& Vauclair, G. 1976, ApJ, 210, 447 (see Figure 6)
\bibitem[Michaud, Richer \& Vick(2011)]{mic11} Michaud, G, Richer, J. \& Vick, M. 2011,
A\&A, 534, 18 (MRV)  
\bibitem[Morton (2000)]{mor00} Morton, D. C. 2000, ApJS, 130, 403 [Erratum: 132, 411, 2001]
\bibitem[Nilsson, et al. (2002b)]{niv02} Nilsson, H., Ivarsson, S., Johansson, S., et al.
2002, A\&A, 381, 1090
\bibitem[Nilsson \& Ivarsson(2008)]{nia08} Nilsson, H. \& Ivarsson, S. 2008, A\&A, 
492, 609
\bibitem[Nilsson, et al. (2008)]{nib08} Nilsson, H., Engstr\"{o}m, Lundberg, H., et al.
2008, Eur Phys. J. D, 49, 13
\bibitem[Nilsson, et al.(2002a)]{nil02} Nilsson, H., Zhang, Z. G., Lundberg, H., et al.
2002, A\&A, 382, 368
\bibitem[Nilsson, et al. (2010)]{nil10} Nilsson, H., Hartman, H., Engstr\"{o}m, L., et al.
2010, A\&A, 511, 16
\bibitem[Oliver \& Hibbert(2010)]{olh10} Oliver, P. \& Hibbert, A., 2010, J. Phys. B,
43, 074013
\bibitem[Ortiz, et al. (2013)]{ort13} Ortiz, M., Arag\'{o}n, C., Aguilera, J. A.,      
et al. 2013, J. Phys. B, 46, 185702
\bibitem[Palmeri, et al. (2001)]{pal01} Palmeri, P., Quinet, P. \& Bi\'{e}mont, \'{E}.
2001, Phys. Scr., 63, 468
\bibitem[Palmeri, et al. (2009)]{pal09} Palmeri, P., Quinet, P., Bi\'{e}mont, \'{E}., 
et al. 2009, J. Phys. B., 42, 165005
\bibitem[Petit, et al. (2014)]{pet14} Petit, P., Louge, T., Th\'{e}ado, S., et al. 2014,
PASP, 126, 469
\bibitem[Proffitt, et al.(1999)]{pro99} Proffitt, C. R., Brage, T., Leckrone, D. S., et al.
1999, ApJ, 512, 942
\bibitem[Qiu, et al. (2001)]{qiu01} Qui, H. M., Zhau, G., Chen, Y. Q., 
et al. 2001, \apj, 548, 953
\bibitem[Quinet (1996)]{qui96} Quinet, P. 1996, Phys. Scr., 54, 483
\bibitem[Quinet, et al. (1999)]{qui99} Quinet, P., Palmeri, P., Bi\'emont, \'{E}, et al.
1999, MNRAS, 307, 934
\bibitem[Quinet, et al. (2006)]{qui06} Quinet, P.,Palmeri, P., Bi\'emont, \'{E}, 
et al. 2006, A\&A, 448, 1207
\bibitem[Quinet, et al. (2008)]{qui08} Quinet, P., Palmeri, P., Fivet, V., et al. 2008,
Phys. Rev. A, 77, 022501
\bibitem[Quinet, et al. (2012)]{qui12} Quinet, P., Bi\'{e}mont, \'{E}., Palmeri, P., et al. 
2012, A\&A, 537, 74

\bibitem[Richer, et al. (2000)]{ric00} Richer, J., Michaud, G. \& Turcotte, S. 2000, 
\apj, 529, 338

\bibitem[Rogerson (1987)]{rog87} Rogerson, J. B., Jr. 1987, ApJS, 63, 369
\bibitem[Ryabchikova, et al. (2006)]{rya06} Ryabchikova, T., Ryabtsev, A., Kochuknov,
O., et al. 2006, A\&A, 456, 329
\bibitem[Ryabchikova, et al. (2015)]{rya15} Ryabchikova, T, Piskunov, N., Kurucz, 
R., et al. 2015, Phys. Scr., 90, 054005
\bibitem[Sadakane, et al. (1988)]{sad88} Sadakane, K., Jugaku, J. \& Takada-Hidai, M. 
1988, PASP, 100, 811
\bibitem[Sadakane(1991)]{sad91} Sadakane, K. 1991, PASP, 103, 355
\bibitem[Sansonetti, Martin \& Young (2005)]{san05}
Sansonetti, J. E., Martin, W. C. \& Young, S. L. 2005, Handbook of Basic 
Atomic Spectroscopic Data (version 1.1.2). [Online] Available: 
http://physics.nist.gov/Handbook  [2015]. National 
Institute of Standards and Technology, Gaithersburg, MD.
\bibitem[Scott, et al. (2015)]{scb15} Scott, P., Grevesse, N., Asplund, M., et al.
2015b, A\&A, 573, A27
\bibitem[Shirai, et al. (2007)]{shi07} Shirai, T., Reader, J., Kramida, A. E. \& Sugar,
J. 2007, J. Phys. Chem. Ref. Data. 36, 509
\bibitem[Silkstrom, et al.(2001)]{sil01} Silkstr\"{o}m, C. M., Pihlemark, H., Litzen, U.,
et al. 2001, J. Phys. B., 34, 477
\bibitem[Silvester, et al. (2012)]{slv12} Silvester, J., Wade, G., Kochukhov, O., et al.
2012, MNRAS, 426, 1003
\bibitem[Smith \& Dworetsky (1993)]{kcs93} Smith, K. C. \& Dworetsky, M. M. in Peculiar
Versus Normal Phenomena in A-Type and Related Stars, ASP Conference Ser., 44, 1993
(ed. M. M. Dworetsky, F. Castelli \& R. Faraggiana). p. 131 
\bibitem[Takeda, et al. (2008)]{tak08} Takeda, Y., Han, I, Kang, D., et al. 2008, 
J. Korean Astron.
Soc., 41, 83
\bibitem[Takeda, et al.(2012)]{tak12} Takeda, Y., Kang, D., Han, I., et al. 2012, 
PASJ, 64, 38
\bibitem[Van Winckel(2003)]{van03} Van Winkel, H. 2003, Ann. Rev. Astron. Ap.,
41, 391 (see Fig. 3, panel 6)
\bibitem[Venn \& Lambert(1990)]{ven90} Venn, K. A. \& Lambert, D. L. 1990,
ApJ, 363, 234    
\bibitem[Wahlgren et al(1993)]{wah93} Wahlgren, G. M., Johansson, Se, Kurucz, R. L.
\& Leckrone, D. S. 1993, Bull. AAS, 25, 1321
\bibitem[Wahlgren, et al. (1997)]{wah97} Wahlgren, G. M., Johansson, S., Litz\'{e}n, U., 
et al. 1997, ApJ, 475, 380
\bibitem[Wahlgren \& Dolk(1998)]{wah98} Wahlgren, G. M. \& Dolk, L. 
1998, Contrib. Astron. Obs. Skalnat\'{e} Pleso, 27, 314
\bibitem[Wahlgren, et al.(2000)]{wah00} Wahlgren, G. M., L., Dolk, L., Kalus, G.,
et al. 2000, ApJ, 539, 908
\bibitem[Wickliffe \& Lawler(1997)]{wic97} Wickliffe, M. E. \& Lawler, J. E. 1997,
JOSA B, 14, 737
\bibitem[Wickliffe, et al.(2000)]{wic00} Wickliffe, M. E., Lawler, J. E. \& Nave, G.
2000, J. Quant. Spec. Rad. Trans., 66, 363
\bibitem[Woodgate et al.(1998)]{woo98} Woodgate, B.~E., Kimble, R. A.,
Bowers, C. W., et al. 1998, PASP, 110, 1183
\bibitem[Xu, et al.(2007)]{xup07} Xu, H. L., Svandberg, S., Quinet, P.,
et al. 2007, JQSRT, 104, 52
\bibitem[Yushchenko \& Gopka(2006)]{yus06} Yushchenko, A. \& Gopka, V. 2006, in
{\it Origin of Matter and Evolution of Galaxies}, ed. S. Kubono, W. Aoki, T. Kajino,
T. Motobayashi \& K. Nomoto (Am. Inst. of Physics), p. 503
\bibitem[Zhang, et al. (2013)]{zha13} Zhang, W., Paleeri, P., Quinet, P., et al.
2013, A\&A, 551, 136 
\bibitem[Zhiguo, et al. (1999)]{zhi99} Zhiguo, Z., Zhongshan, L. \& Zhankui, J. 1999,
Eur. Phys. Journal D 7,499
\end{thebibliography}
\end{document}